\documentclass[twocolumn,10pt]{tsfp}
\usepackage{flushend}
\usepackage{graphicx}
\usepackage[authoryear,round]{natbib}
\usepackage{fancyhdr}
 
\usepackage{paralist}
\usepackage{wrapfig}
\usepackage{enumitem}
\usepackage[export]{adjustbox}

\def\gf1{0.3}

\title{Continuous Eddy Simulation (CES): Conceptual Approach and Applications}

\renewcommand{\expauthors}{0.95}

\author{Stefan Heinz
    \affiliation{
	Department of Mathematics \& Statistics\\
	University of Wyoming\\
	1000 E. Univ. Ave., Laramie, WY 82071, USA\\
    heinz@uwyo.edu
    }	
		}

\author{Adeyemi Fagbade
    \affiliation{
	Department of Mathematics \& Statistics\\
	University of Wyoming\\
	1000 E. Univ. Ave., Laramie, WY 82071, USA\\
    afagbade@uwyo.edu
    }
		}

\begin{document}

\maketitle   
\thispagestyle{fancy}

\fontsize{9}{11}\selectfont

\section*{ABSTRACT}

The simulation of high Reynolds number ($Re$) separated turbulent flows faces significant problems for decades: large eddy simulation (LES) is computationally too expensive, and Reynolds-averaged Navier-Stokes (RANS) methods and hybrid RANS-LES methods often provide unreliable results. This has serious consequences, we are currently unable to reliably predict very high $Re$ regimes, which hampers applications and our understanding of turbulence structures. 
The paper reports the advantages of a strict mathematical approach, continuous eddy simulation (CES), to derive partially resolving turbulence models. In contrast to popular hybrid RANS-LES, this minimal error approach includes a dynamic modification of the turbulence model in response to the actual flow resolution: the model can increase (decrease) its contribution to the simulation in dependence of a low (high) flow resolution. 
This property is the essential requirement to seamlessly cover RANS and LES regimes. 
The CES modeling approach offers essential advantages regarding its functionality: basically, it is independent of a variety of simulation settings applied in popular hybrid RANS-LES to improve the model performance. In addition, the CES computational cost can be below the cost of other hybrid RANS-LES and LES by orders of magnitude. Essential simulation performance advantages of CES simulations are described here with respect to three complex flow applications: periodic hill flows at high Reynolds number~\citep{PF20}, the NASA wall-mounted hump flow~\citep{SeilPack}, and the Bachalo \& Johnson axisymmetric transonic bump 
flow~\citep{Bachalo-86, Lynch-20}.

\section*{INTRODUCTION}
\label{sec:intro}

Turbulent flows of practical relevance are often characterized by a high $Re$ and (depending on the flow geometry) flow separation. It is well-known that reliable and efficient simulations of separated high $Re$ turbulent flows face significant problems. Reynolds-averaged Navier-Stokes (RANS) methods are known to be often unreliable. Large eddy simulation (LES) is aiming at a realistic simulation of instantaneous turbulence, but LES is limited to not too high $Re$ flow simulations because of significant computational cost requirements.

The most appropriate balance between RANS and LES approaches is the design of hybrid RANS-LES methods~\citep{JPAS-20, Menter-21}. A large variety of such hybrid RANS-LES was introduced so far. 
The problem of these hybrid RANS-LES is their limited reliability: there is a lack of proven predictive power, all such hybrid RANS-LES results need validation, which is often hardly possible. 
For example, a usual problem of popular hybrid RANS-LES, wall-modeled LES (WMLES)~\citep{Larsson-16} and detached eddy simulation (DES) methods~\citep{Mocket-12}, is the significant uncertainty of predictions depending on adjustable model settings. Such predictions depend on different (equilibrium, nonequilibrium) wall models applied, definitions of regions where different models and grids are applied, different mesh distributions, and set-up options to manage the data transfer between such different flow regions~\citep{JPAS-20}. 

An explanation for the problems faced by popular hybrid RANS-LES provides the concept of these methods to generate conditions which enable the model to produce as much as possible resolved motion. This idea does not ensure at all a physically meaningful simulation. The best example is LES performed on coarse grids. Although there is a lot of resolved motion, such simulations are known to usually provide unphysical results. It is plausible that the uncontrolled balance between modeled motion represented by the model  and produced resolved motion implies the sensitive dependence of model results on changes of simulation settings (see the discussion in the preceding paragraph). 
The most relevant problem is that usually applied hybrid RANS-LES do not enable a seamless transition between RANS and LES regimes. For example, simulations of the effect of increasing $Re$ using the same grid face a decrease of the amount of resolved motion. The model needs to compensate this loss of total kinetic energy by increasing its contribution to the simulation. In general, a functional RANS-LES swing requires that the model contribution to the simulation is relatively low (high) if the flow resolution is high (low). The latter requires that the model receives information about the amount of resolved motion, which is not the case in regard to usually applied hybrid RANS-LES. 

These issues lead to the question of which more appropriate simulation concepts can be applied to overcome these serious problems. The most promising approach is the design of a computational method that minimizes the hybridization error: the expectation is that a minimal error requires an appropriate balance of resolved and modeled motion. Such methods 
[continuous eddy simulation (CES) methods] were introduced recently: see Refs.~\citep{PF19, PF20, PFCES-21, PF22, ApplSci24-CES, Fluids24-CES}. The motivation for presenting this paper is to describe the main characteristics of CES methods in conjunction with their ability to overcome the issues of popular hybrid RANS-LES.

\begin{table*}[t]
		\centering
\scalebox{0.83}{
\begin{tabular}{@{} l l @{}}  

\hline\hline  \vspace{0.05cm} 
\rule{0pt}{4.9ex} \hspace{-0.25cm} \(\displaystyle  \frac{ D k }{ D t} =P-\epsilon+D_k,   ~~~~~ \frac{ D \omega}{ D t} =C_{\omega_1} \omega^2 \Big( \frac{P}{\epsilon}-\beta^{*} \Big) + D_{\omega}  \), ~~~ 
\(\displaystyle D_k=\frac{\partial}{\partial x_j} \nu_t^* \frac{\partial k}{\partial x_j}\),~~~
\(\displaystyle D_\omega=\frac{\partial}{\partial x_j} \frac{\nu_t^*}{\sigma_\omega} \frac{\partial \omega}{\partial x_j} \)
 \hspace{1.9cm}(KOS)
 \\[2.4ex]  

\rule{0pt}{1.ex} \hspace{-0.25cm} \(\displaystyle  \frac{ D k }{ D t} =P-\psi_{\beta}  \epsilon+D_k,   ~~ \frac{ D \omega}{ D t} =C_{\omega_1} \omega^2 \Big( \frac{P}{\epsilon}-\beta \Big) +D_\omega  \), ~~~~~$\beta^{*}=1+ \beta-\psi_{\beta}$ 
\hspace{4.85cm}(KOK)

\\[2.7ex]  
\hline   

\rule{0pt}{5.ex} \hspace{-0.2cm} \textbullet\ Analysis option $\mathcal{O}_1$ [with $Dk/D t, D \omega/Dt$, $\nu_t^*=\nu_{t,tot}$]:~~~~ 
 
\bigg[\(\displaystyle \frac{\delta (D k/D t)}{D k/D t}=\frac{\delta D_{k}}{D_{k}}=\frac{\delta k}{k} \),~ 
\(\displaystyle \frac{\delta (D \omega/D t)}{D \omega/D t}=\frac{\delta D_{\omega}}{D_{\omega}}=\frac{\delta \omega}{\omega}\bigg] \) 
 \\[2.2ex]

\rule{0pt}{4.2ex} \hspace{0.cm} 
~\(\displaystyle 
 \lambda_1=C_{\omega_1} \omega^2 \Big( \frac{P}{\epsilon}-\beta_1^* \Big) + D_{\omega} -\frac{D \omega}{Dt}\),  
~ \(\displaystyle 
\delta \Big( \frac{\lambda_1}{\omega} \Big) = \frac{C_{\omega_1}}{\tau}(\beta_1^{*}-1) \Big[ 
\frac{\delta \tau}{\tau} -\frac{\delta \beta_1^{*}}{\beta_1^{*}-1} \Big] \), 
~ \(\displaystyle \int_{\beta}^{\beta_1^{*}} \!\!\!\! \frac{dx}{x-1} =\! \int_{\tau_{tot}}^{\tau} \!\! \frac{dy }{y}  \),
~~ \(\displaystyle \frac{\beta_1^{*}-1}{\beta-1}=\tau_{+} \)
 \\[2.2ex]

\rule{0pt}{3.5ex} \hspace{-0.2cm} \textbullet\ Analysis option $\mathcal{O}_2$ [without $Dk/D t, D \omega/Dt$, $\nu_t^*=\nu_t$]:~~~~  \big[$\delta D_{k}/D_{k}=3\delta k/k-\delta \epsilon/\epsilon$,~ $\delta D_{\omega}/D_{\omega}=\delta k/k$\big] ~~~\\[-0.5ex]

\rule{0pt}{5.4ex} \hspace{0.cm} 
~\(\displaystyle 
 \lambda_2=C_{\omega_1} \omega^2 \Big( \frac{P}{\epsilon}-\beta_2^* \Big) + D_{\omega} \),  ~
~\(\displaystyle 
\delta \Big( \frac{\lambda_2}{k} \Big) =\frac{C_{\omega_1}}{L^2}(\beta_2^{*}-1) \Big[ 
\frac{\delta L^2}{L^2}-\frac{\delta \beta_2^{*}}{\beta_2^{*}-1} \Big] \),
~~ \(\displaystyle \int_{\beta}^{\beta_2^{*}} \!\!\!\! \frac{dx}{x-1} =\! \int_{L^2_{tot}}^{L^2} \!\! \frac{dy }{y}  \),
~~~ \(\displaystyle \frac{\beta_2^{*}-1}{\beta-1}=L_{+}^{2} \)
\\[3.ex]

\hline\hline  
\end{tabular} 
}  
\vspace{-0.cm} \caption{Minimal error $k-\omega$ models: both KOS and KOK hybridizations are considered in analysis options $\mathcal{O}_1$, $\mathcal{O}_2$ depending on $\nu_t^*$. Model errors $\lambda$, first variations, and resulting mode controls $\beta^{*}$ are provided. Variations applied are given in brackets.
} 		
\label{tab:CES-KOS}
\vspace{-0.cm}
\end{table*}

\section*{MINIMAL ERROR SIMULATION METHODS}
\label{sec:problem}

Incompressible flow is considered for simplicity, corresponding compressible formulations can be found elsewhere. The incompressible continuity equation $\partial\widetilde{U}_{i} / \partial x_{i}=0$ 
and momentum equation are considered, 
\vspace{-0.4cm}
\begin{equation}
\frac{D\widetilde{U}_{i}}{Dt}=-\frac{\partial  (\widetilde{p}/\rho+2k/3) }{\partial x_{i}}+2 \frac{\partial (\nu+\nu_{t}) \widetilde{S}_{ik}}{\partial x_{k}}.    
\label{eq:momentum}
\vspace{-0.4cm}
\end{equation}
Here, $D/ D t=\partial/\partial t+\widetilde{U}_{k}\partial/\partial x_{k}$ denotes the filtered Lagrangian time derivative and the sum convention is used throughout this paper. $\widetilde{U}_{i}$ refers to the $i^{th}$ component of the spatially filtered velocity. We have here the filtered pressure $\widetilde{p}$, $\rho $ is the constant mass density, $k$ is the modeled energy, $\nu$ is the constant kinematic viscosity, and 
$\widetilde{S}_{ij}=(\partial \widetilde{U}_{i}/\partial x_{j}+\partial \widetilde{U}_{j}/\partial x_{i})/2$ is the rate-of-strain tensor.
The modeled viscosity is given by $\nu_{t}=C_{\mu} k^{1/2}L$. Here, $C_{\mu}$ is a model parameter with standard value $C_{\mu}=0.09$, and $L$ is a characteristic length scale. $L$ can be calculated in different ways using $L=k^{3/2}/\epsilon=k^{1/2}\tau=k^{1/2}/\omega$, where the dissipation rate $\epsilon=k/\tau$ of modeled kinetic energy, the dissipation time scale $\tau$, and the turbulence frequency $\omega=1/\tau$ are involved. 

The minimal error approach can be applied in conjunction with a variety of turbulence models. Table~\ref{tab:CES-KOS} shows its application in regard to the $k-\omega$ model. $P=\nu_{t} S^{2}$ is the production of $k$, where $S=(2\widetilde{S}_{mn} \widetilde{S}_{nm})^{1/2}$. We have here $C_{\omega_1}=0.49$, and $\sigma_\omega=1.8$. 
In RANS, $\beta^{*}$ is considered to be constant. Here, $\beta^{*}$ is considered to be an undetermined parameter that needs to be chosen to minimize the hybridization error. 
The abbreviations KOS and KOK refer to the consideration of the $k-\omega$ model where the hybridization is accomplished in the scale equation (involving $\beta^{*}$ that needs to be determined) or $k$-equation (involving $\psi_{\beta}$ that needs to be determined), respectively. Both approaches can provide equivalent results as long as the coefficient relation $\beta^{*}=1+ \beta-\psi_{\beta}$ is taken into account. In regard to both KOS and KOK models there are two analysis options. Option $O_1$ is an exact hybridization where total viscosities need to be applied in turbulent transport terms. Option $O_2$ is an hybridization where the usual model viscosities are involved in turbulent transport terms; an approximation is involved by the neglect of substantial derivatives in regard to the model coefficient calculation. 

\begin{figure}[b]
		\begin{center}
			\begin{tabular}[c]{cc}
				\hspace{-0.35cm}
				{\includegraphics[width=0.25\textwidth]{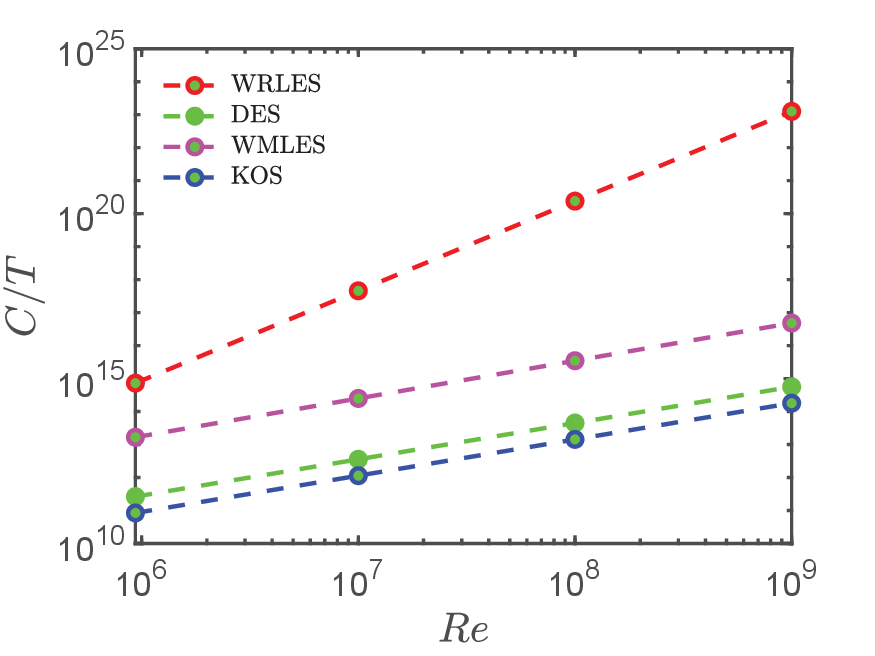}}\hspace{-0.35cm}
			{\includegraphics[width=0.25\textwidth]{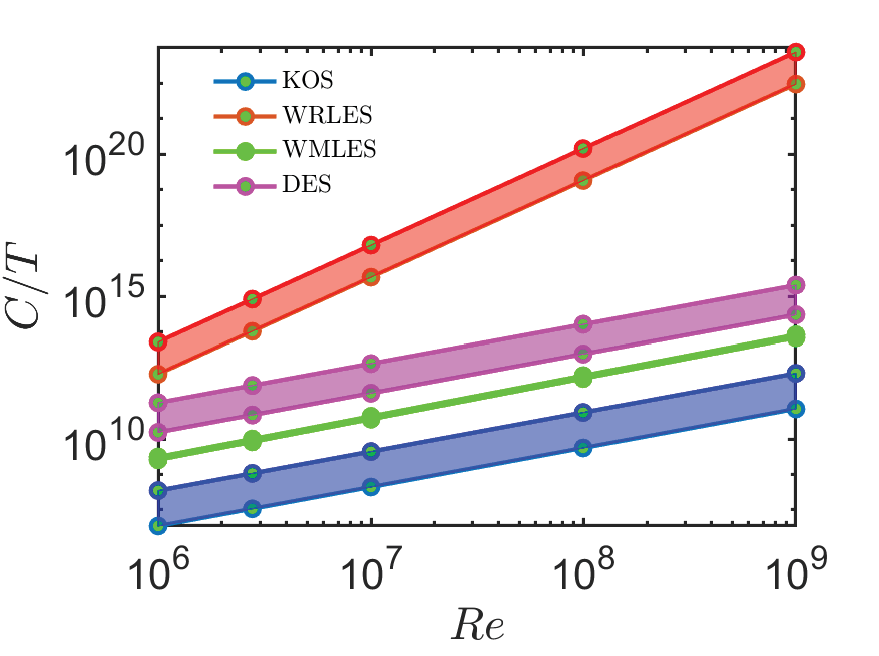}}\\
			\end{tabular}
		\end{center}
		\vspace{-0.5cm}
		\caption{Cost scalings of CES vs. other methods: NASA wall-mounted hump flow (left) and the Bachalo \& Johnson axisymmetric transonic bump flow (right). }
		\label{fig:cost-SH}
		\vspace{0.3cm}
	\end{figure}

The structure of CES methods differs significantly from popular RANS-LES methods due to the involvement of resolution indicators like $0 \leq L_{+} \leq 1$ in model equations. The latter measures the degree of flow resolution.  $L_+=L/L_{tot}$ is defined in similarity to the modeled-to-total kinetic energy ratio $k_+=k/k_{tot}$, where $L$ refers to the modeled length scale contribution and $L_{tot}$ refers to the total length scale contribution. In particular, $L_{+}\approx 1$ represents an almost completely modeled (RANS) regime, whereas $L_{+}\approx0$ represents an almost completely resolved (LES) regime. Most importantly, in contrast to usually applied hybrid RANS-LES the model is informed about the actual resolution in this way, and the model is able to respond to resolution variations implied by changes of the model coefficient $\beta^{*}$. For example, a higher resolution ($L_{+}$ becomes smaller) decreases $\beta^{*}$. Thus, there is less dissipation of $\omega$, $\omega$ increases which decreases the model viscosity $\nu_{t}=C_{\mu} k/\omega$. To minimize the error of a hybrid method means, therefore, to minimize the uncontrolled coexistence of resolved and modeled motion seen in popular hybrid RANS-LES.

\section*{CES FUNCTIONALITY AND COST FEATURES}

It is worth noting that the CES approach reveal essential differences to usually applied wall-resolved LES (WRLES), WMLES, and DES. 
WRLES requires the  use sufficiently fine grids, and it is often unclear whether LES resolution requirements are satisfied. Hybrid RANS-LES like WMLES methods are known to depend on simulation settings, the use of different (equilibrium or nonequilibrium) wall models, definitions of regions where different models
and grids are applied, different mesh distributions, and setup options to manage the information exchange between such different flow regions. DES depends similarly on simulation settings, the results depend on the concrete model applied and the definition of differently treated simulation zones. Both DES and WMLES are known to depend on the mesh organization using the same number of grid points. In contrast, CES methods are independent of such functionality requirements, the model can be used as is. In particular, CES can be expected to enable reliable predictions under conditions where validation data are unavailable.

Also in regard to the computational cost of CES methods, there are essential differences to usually applied methods~\citep{PF20, ApplSci24-CES, Fluids24-CES}. The simulation cost are specified by $C= N N_{t} = TN/\Delta t
$. 
Here, $N$ is the number of grid points applied, $N_{t}$ is number of time steps performed, $T = N_{t}\Delta t$  refers to the constant total physical simulation time, and $\Delta t$ is the prescribed simulation time-step.
$N$ and $\Delta t$ are known to vary with $Re$ according to $N= \alpha_{1} (Re/Re_0)^{\beta_{1}}$, $\Delta t= \alpha_{2} (Re/Re_0)^{-\beta_{2}}$, where $\alpha_{1}$, $\alpha_{2}$, $\beta_{1}$, and $\beta_{2}$ are 
constants~\citep{Texas-15, PF16}. Here, $Re_0$ is used as normalization. Implications of simulations of NASA wall-mounted hump flow and the Bachalo \& Johnson axisymmetric transonic bump flow are presented in Fig.~\ref{fig:cost-SH}. As it may be seen, the simulation cost of CES are well below the cost of other methods, in particular, CES applications can be by orders of magnitude cheaper than other methods. 

\begin{figure}[b]
    \includegraphics[width=0.45\textwidth]{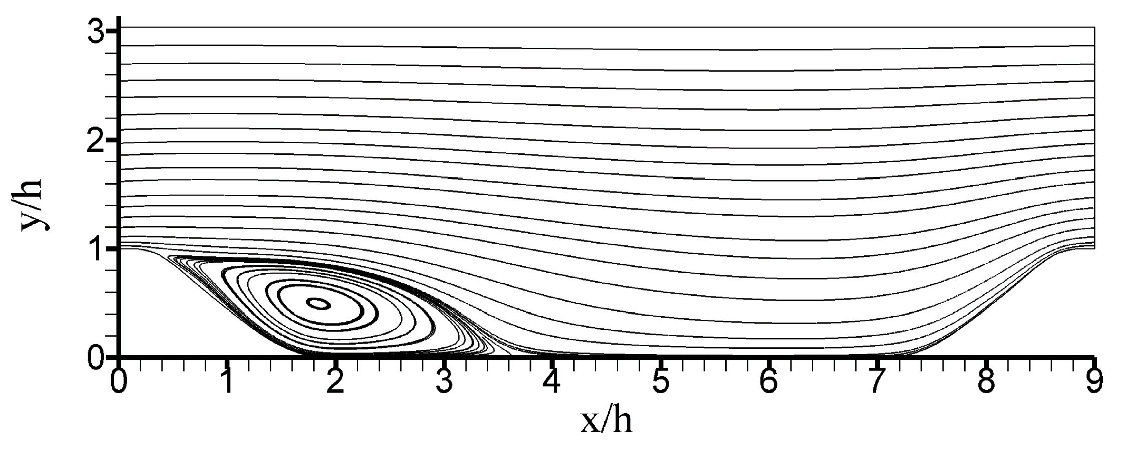}
    \caption{
       Velocity streamlines seen in periodic hill flows: results obtained by continuous eddy simulation at $Re=37,000$. Reprinted with permission from Ref.~\citep{PF20}. Copyright 2020 AIP Publishing.
    } \label{fig:streamlines}
	\vspace{-0.cm}
\end{figure}

\begin{figure*}[ht!]
\begin{center}
\begin{tabular}[c]{cccccc}
\hspace{-0.4cm} {\includegraphics[width=0.32\textwidth]{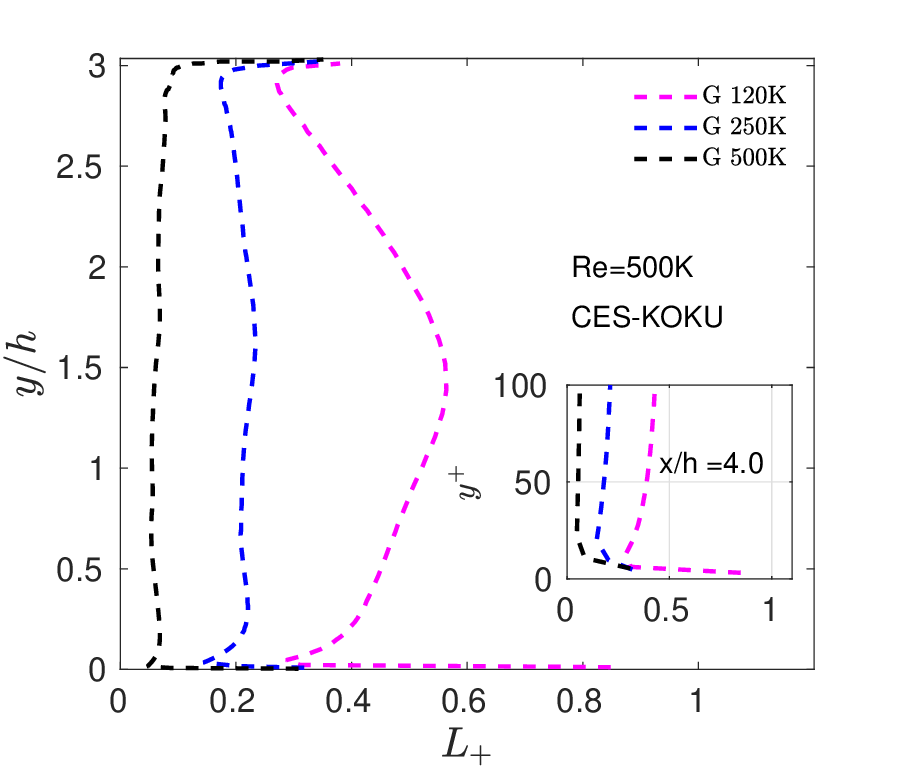}  }&
\hspace{-0.4cm} {\includegraphics[width=0.32\textwidth]{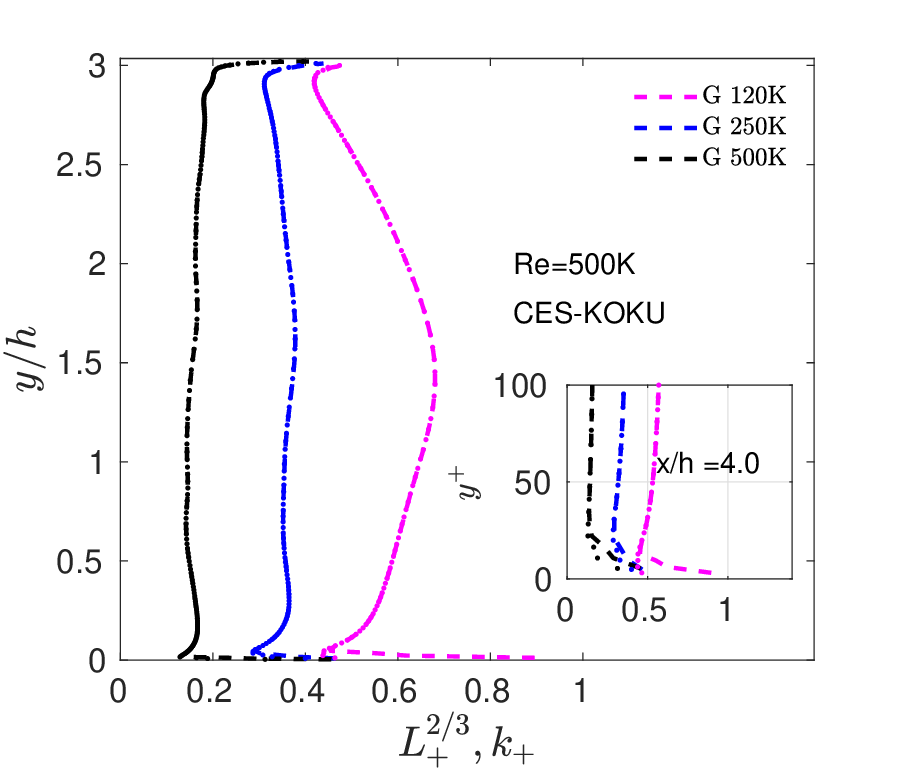}  }&
\hspace{-0.4cm} {\includegraphics[width=0.32\textwidth]{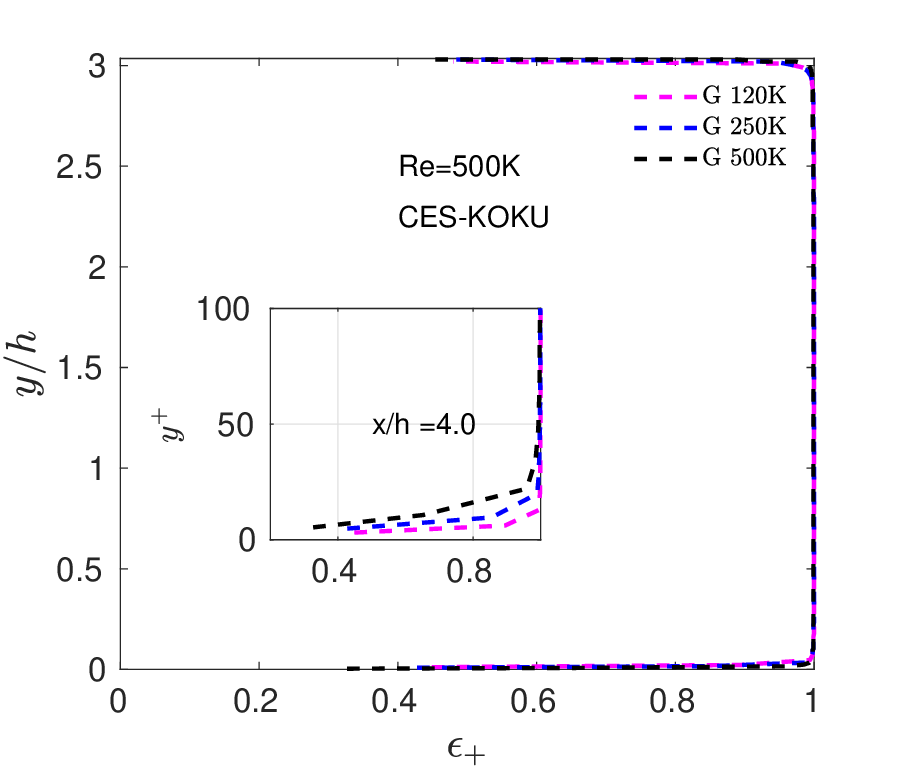}  }\\
\hspace{-0.4cm} {\includegraphics[width=0.32\textwidth]{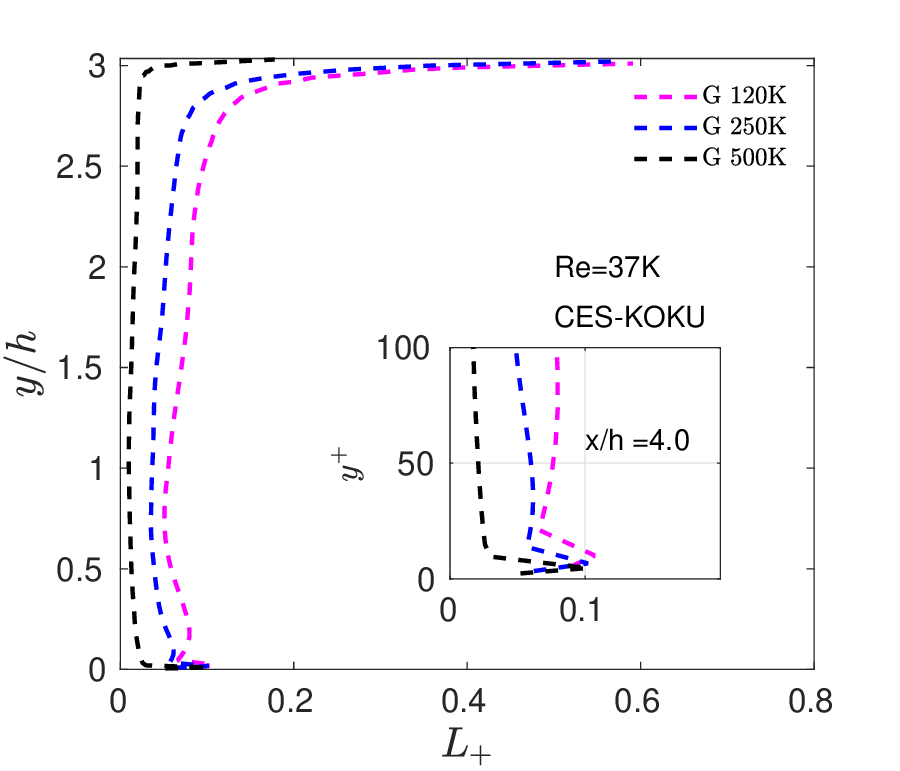}  }&
\hspace{-0.4cm} {\includegraphics[width=0.32\textwidth]{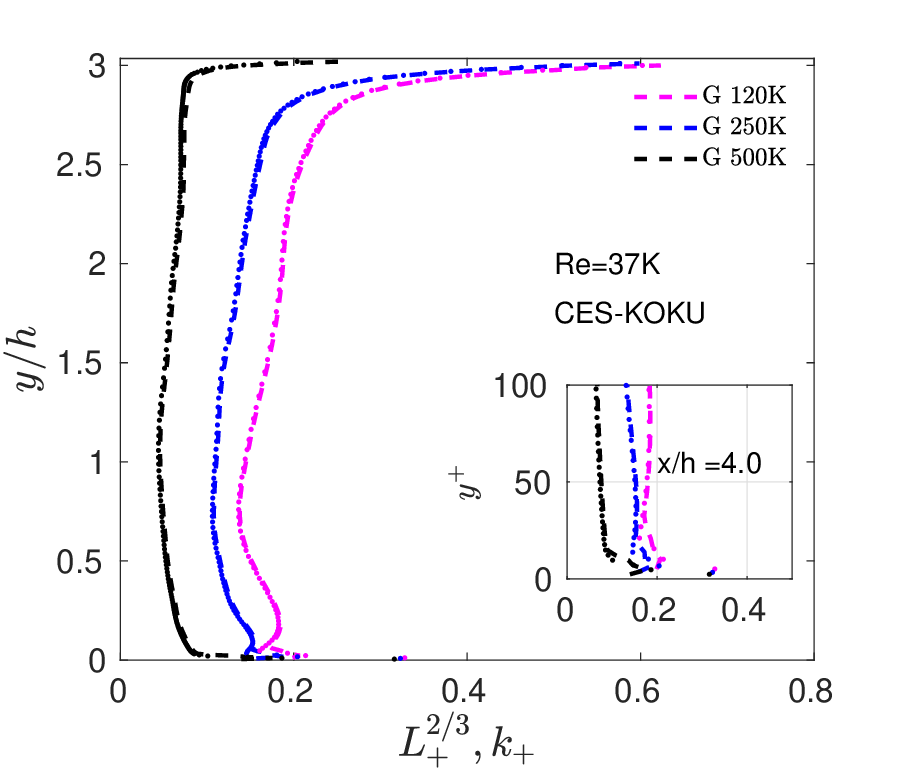}  }&
\hspace{-0.4cm} {\includegraphics[width=0.32\textwidth]{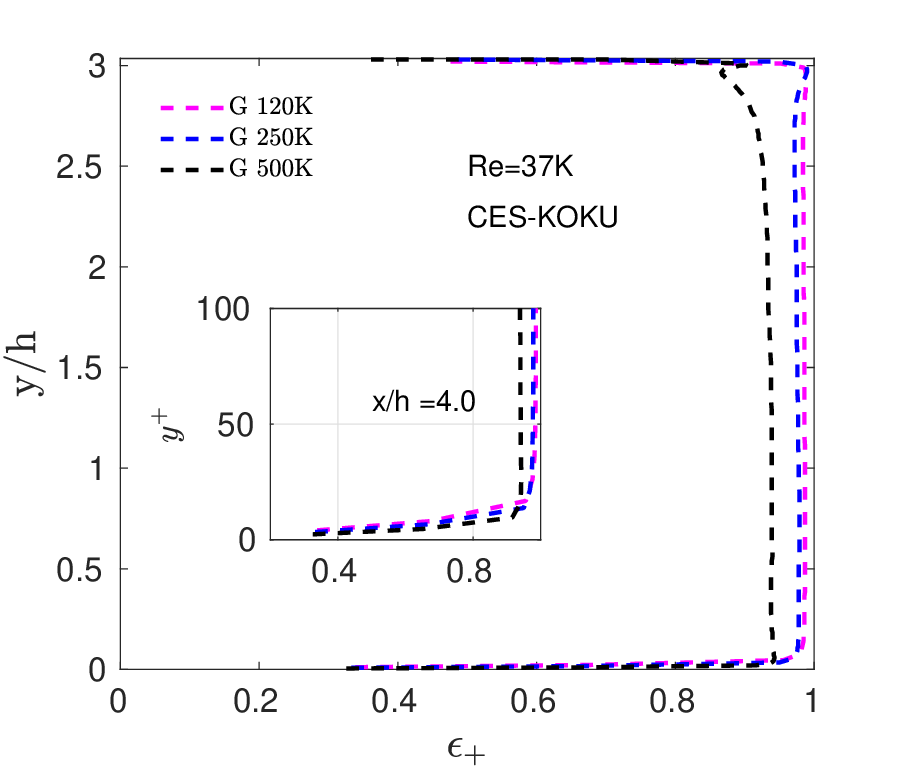}  }\\
   \end{tabular}
\end{center}
\vspace{-.3cm}  \linespread{1.}
\caption{CES-KOKU results for the resolution indicators $L_+$, $k_+$, and $\epsilon_+$: Results are shown for  $Re=500K$ and $Re=37K$, respectively, on $G_{500}$, $G_{250}$, $G_{120}$. 
The insets show profiles close to the lower wall in $y_{+}$ scaling. Reprinted with permission from Ref.~\citep{PF20}. Copyright 2020 AIP Publishing.}
\label{fig:allresolution}
\vspace{0.cm}
\end{figure*}

\begin{figure*}[t]
\begin{center}
\begin{tabular}[c]{cccccc}
\hspace{-0.4cm} {\includegraphics[width=\gf1\textwidth]{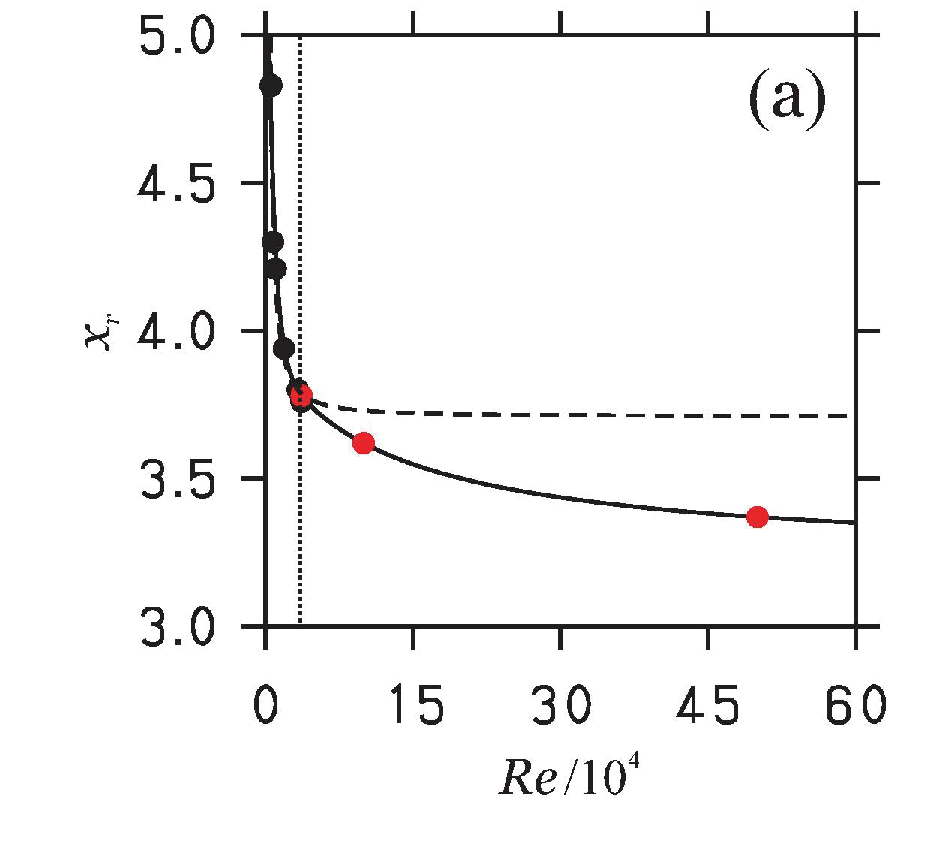}  }
\hspace{-0.4cm} {\includegraphics[width=\gf1\textwidth]{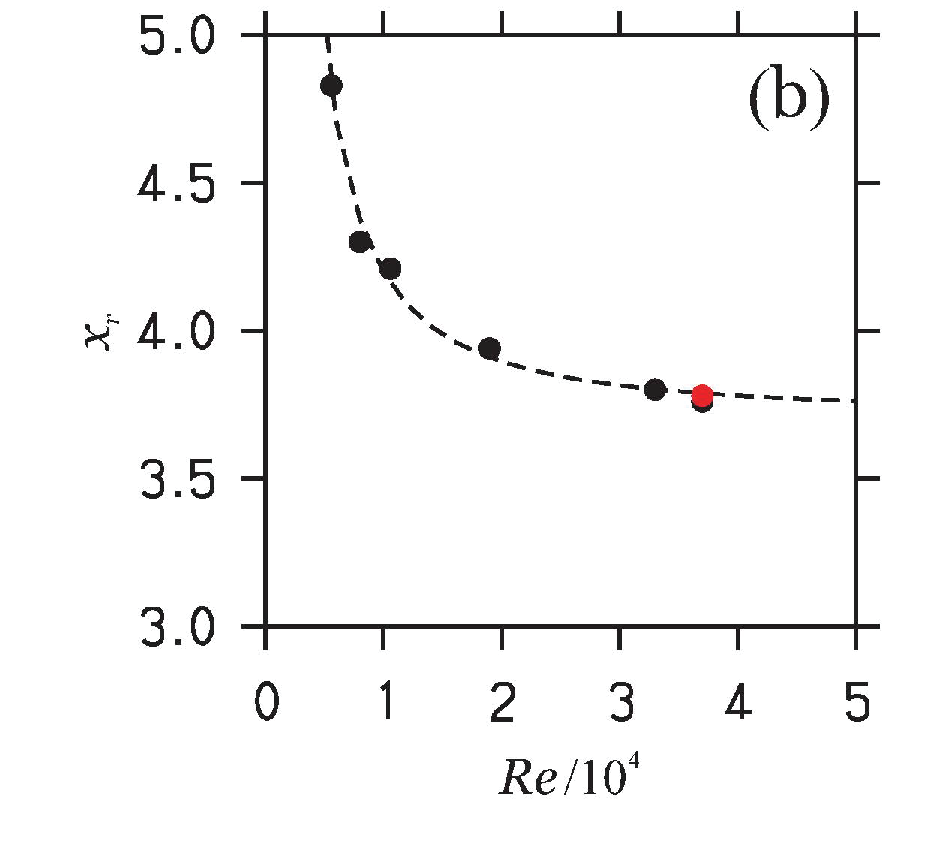}  } 
   \end{tabular}
\end{center}
\linespread{1.} \vspace{-.3in}
\caption{
Reattachment point predictions $x_{r}$ versus $Re$: experimental (black dots)~\citep{Rapp2011, Kaehler-16} and CES results (red dots).  The black line in (a) shows $x_r=3.23[1+15.1 \times 10^{4}/Re][1+0.5e^{-1.5 \times 10^{-4} Re}]/[1+12.4 \times 10^{4}/Re$, 
the dashed line shows  $0.49(Re/10^{4})^{-1.4}+3.71$ of K{\"a}hler et al.~\citep{Kaehler-16}. The vertical dotted line shows the range of previously available results. Figure (b) shows the zoomed-in curve fit of~\citep{Kaehler-16} compared to previous experimental and $Re=37K$ CES results. Reprinted with permission from Ref.~\citep{PF20}. Copyright 2020 AIP Publishing.
}
\label{fig:HighRe}
\vspace{-0.cm}
\end{figure*}

\section*{PERIODIC HILL FLOW SIMULATIONS}

One of the applications of CES methods is the simulation of periodic hill flows as illustrated in Fig.~\ref{fig:streamlines}~\citep{PF20}. This flow is a channel flow involving periodic restrictions. 
This flow, which is used a lot for the evaluation of turbulence models~\citep{JPAS-20}, involves features such as separation, recirculation, and natural reattachment~\citep{Rapp2011, Kaehler-16}. A thorough evaluation of the performance of CES methods in regard to simulating periodic hill flows at the highest $Re=37,000$ for which experimental data for model evaluation are still available can be found elsewhere~\citep{PF20}. The CES-KOKU CES variant was applied~\citep{PF20}. 

An interesting question concerns the spatial uniformity of variations of the distribution of the resolution indicators 
$L_+$, $k_+$, and $\epsilon_+=\epsilon/\epsilon_{tot}$, which is certainly a desired feature to avoid imbalances of resolved and modeled motions. 
Fig.~\ref{fig:allresolution} demonstrates the ability of the hybrid model to produce almost uniform distributions of 
$L_+$ over most of the domain. The most noticeable deviation from this trend is given for 
$Re=500K$ ($G_{120}$). 
Another question concerns the uniformity of resolution indicator variations in response to grid and $Re$ variations. The question is whether there are indications of discontinuities in this regard, which would lead to questions about the applicability of CES methods for very high $Re$ flows using rather coarse grids. The desired  increase of variables like $L_+$ and $k_+$ due to coarser grids and increased $Re$ requires a corresponding reduction of resolved motion (fluctuations), which requires a stable functioning of the fluctuation generation mechanism. 
Fig.~\ref{fig:allresolution} clearly demonstrates that there is a uniform response of $L_+$ to both grid coarsening and increased $Re$, including the near-wall region behavior (see, e.g., the insets showing profiles close to the lower wall in $y_{+}$ scaling).

A representative example of advantages is given in Fig.~\ref{fig:HighRe}. 
The question about the asymptotic flow structure matters to our understanding to see whether there are asymptotically stable regimes of wall-bounded turbulent flows involving flow separation. 
This question relates to our understanding of which geometric conditions enable an asymptotically stable flow configuration~\citep{Fluids22-Mem}, what is the corresponding concrete flow structure, and which $Re$ are needed to accomplish an asymptotic flow structure. 
Figure~\ref{fig:HighRe} shows reattachment point predictions of experiments and CES results depending on $Re$. Available experimental results support the curve fit $0.49(Re/10^{4})^{-1.4}+3.71$ of~\citet{Kaehler-16}. However, the availability of CES predictions reveals the unphysical behavior of this fit for high $Re$, it cannot be expected that the $Re$ trend ends right after the available data range. The curve fit derived from CES predictions, $x_r=3.23[1+15.1 \times 10^{4}/Re][1+0.5e^{-1.5 \times 10^{-4} Re}]/[1+12.4 \times 10^{4}/Re$, provides a more plausible explanation of how the asymptotic $Re$ regime is reached. 

	\begin{figure*}[htbp]
	\centering
	\begin{minipage}{0.4\textwidth}
		\centering
		\includegraphics[width=.7\linewidth]{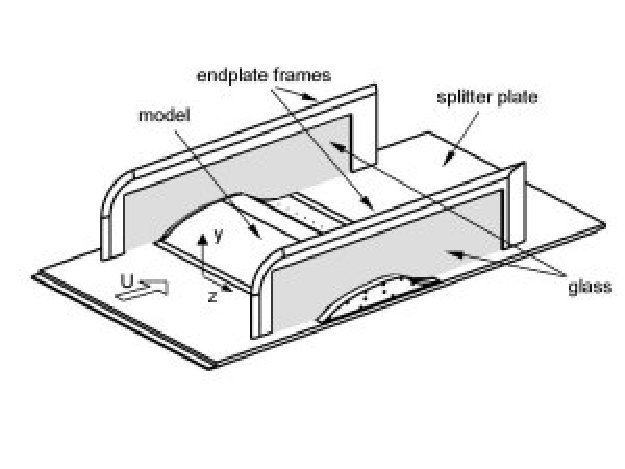}
		\vspace{-0.cm}
	\end{minipage}%
	\begin{minipage}{0.65\textwidth}
		\centering
		\vspace{-0.9cm} \includegraphics[width=0.9\linewidth]{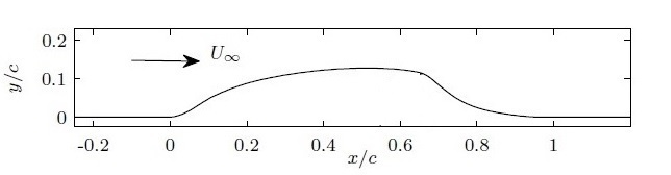}
		\vspace{-0.8cm}
	\end{minipage}
	\vspace{-0.8cm}
	\caption{Wall-mounted hump geometry. Left:  Experimental setup~\citep{SeilPack}; right: 2-D Computational layout.}
	\label{fig:figure1}
\end{figure*}

\begin{figure*}[htbp]
	\centering
	\begin{tabular}{cc}
		\includegraphics[width=0.295\textwidth]{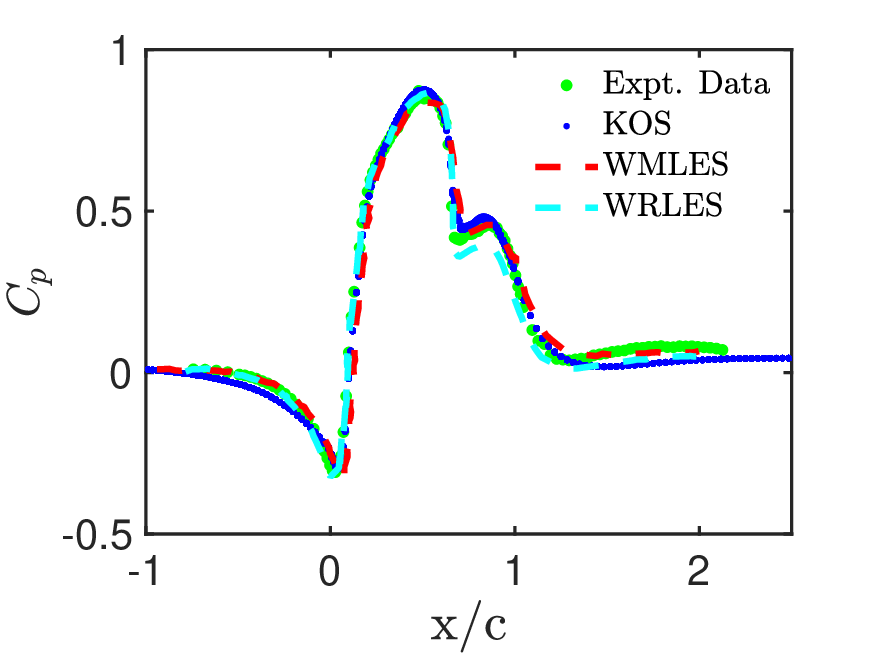} \hspace{-0.35cm}
		\includegraphics[width=0.295\textwidth]{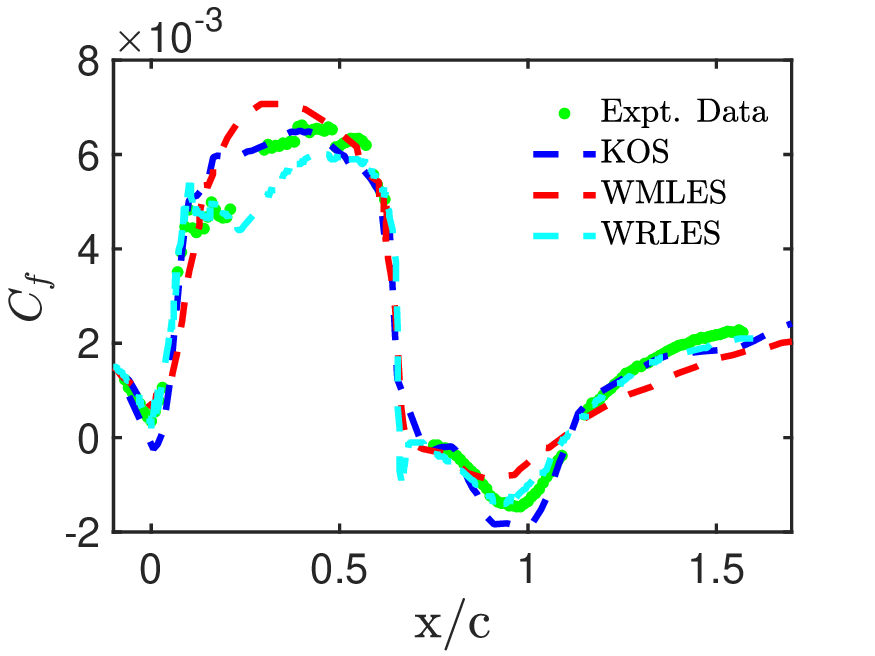} \hspace{-0.35cm}\\
	\end{tabular}
	\vspace{-.15in}
	\caption{CES-KOS, WMLES~\cite{IyerMk}, and WRLES~\cite{UzunMalik2, UzunMalik} simulation results on the $G_4$ grid at $Re = 936K$: Pressure and skin-friction coefficients.
	}
	\label{fig:FCLpppAB4}
\end{figure*}

\section*{NASA HUMP FLOW SIMULATIONS}

\citet{SeilPack} developed the wall-mounted hump model to investigate unsteady flow separation, reattachment, and flow control at a high Reynolds number $Re =c \rho_{ref}U_{ref}/\mu \approx 936K$ based on the chord length $c$ and freestream velocity $U_{ref}$. Here, $\mu$ is the dynamic viscosity and the abbreviation $ref$ indicates the reference freestream conditions, which are determined at the axial point $x/c = -2.14$.
The model reflects the upper surface of a 20-thick Glauert-Goldschmied airfoil that was originally designed for flow control purposes in the early twentieth century.
As a benchmark for comparison, we used the experiment conducted by Greenblatt et al. \citep{Greenblatt} without flow control. This benchmark case has been extensively documented on the NASA Langley Research Center's Turbulence Modeling Resource webpage and has been widely used for evaluating different turbulence modeling techniques, as discussed in the 2004 CFD Validation Workshop. We see in Figure~\ref{fig:figure1} a strongly convex region just before the trailing edge, which induces flow separation.

A representative example of advantages is given in Fig.~\ref{fig:FCLpppAB4}. 
This figure shows that that  all methods involved in this comparison show a reasonable agreement with the experimental pressure coefficient profiles. The predictions from WRLES match the experimental measurement profile downstream and the model is capable of mimicking the dominant features of the flow. However, within the reattachment region, the second wall pressure peak is underpredicted by WRLES compared to CES-KOS and WMLES.
Figure \ref{fig:FCLpppAB4} also shows the mean skin friction coefficient  obtained by CES-KOS, WMLES, and WRLES simulations, demonstrating their agreement with experimental values.  
In the separation zone, from $0\leq x/c \leq 0.65$, WRLES underpredicts the skin friction coefficient, while WMLES overestimates the actual peak. In regard to post-reattachment, however, the $C_{f}$ profiles of WRLES and CES-KOS match relatively well, despite using different frameworks, mesh sizes, and grid resolutions.

	\begin{figure*}[htbp]
		\centering
		\includegraphics[width=0.87\textwidth]{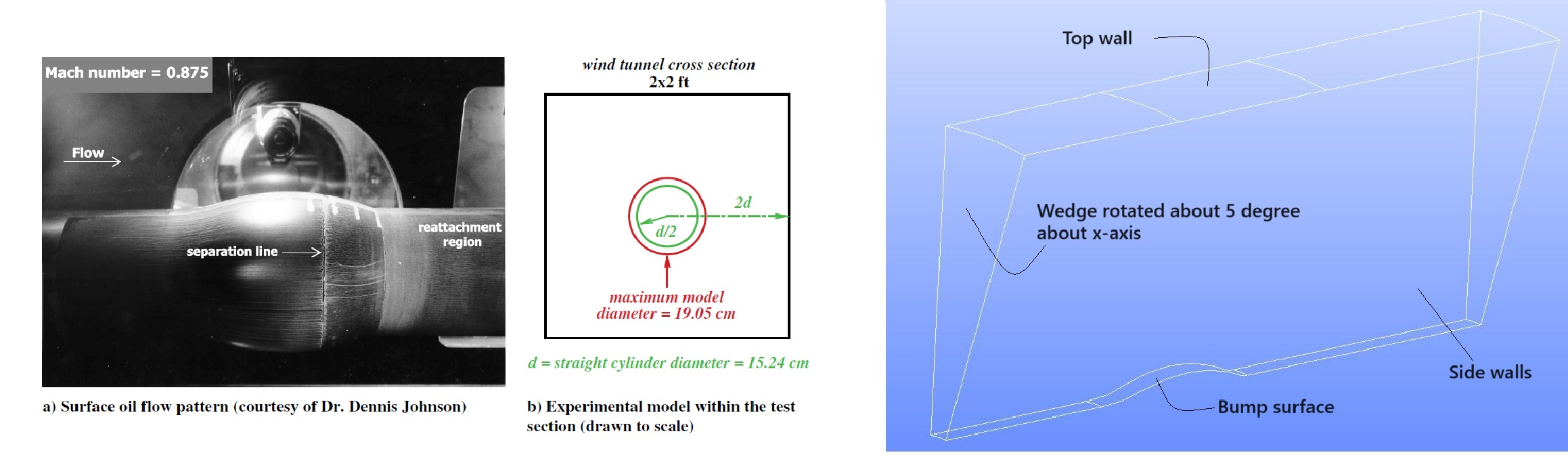}
		\caption{Axisymmetric transonic bump geometry: Experimental and computational configuration~\citep{Xiang_et_all, UZUNMALIK-19}.}
		\label{fig:Simulation set-upA1}
		\vspace{-0.1cm}
	\end{figure*}

\section*{BACHALO  \& JOHNSON AXISYMMETRIC TRANSONIC BUMP FLOW SIMULATIONS}

Figure~\ref{fig:Simulation set-upA1} shows a schematic diagram of the experimental configuration and the computational domain for the axisymmetric transonic bump considered~\citep{Bachalo-86, Lynch-20} along with the applied boundary conditions. 
This case pertains to shock-triggered boundary layer separation induced by an axially-symmetric bump mounted on a slim spherical cylinder, which extends 61cm upstream. The case reflects the upper surface of a transonic wing.  
It is characterized by a Mach number ($M_{\infty}$) of $0.875$ and a Reynolds number ($Re$) of $2.763$ M relative to the airfoil's chord length $c$.

\begin{figure*}[t]	
	\begin{center}
		\begin{tabular}[c]{ccc}
			\hspace{-0.55cm} 
			{\includegraphics[width=0.25\textwidth]{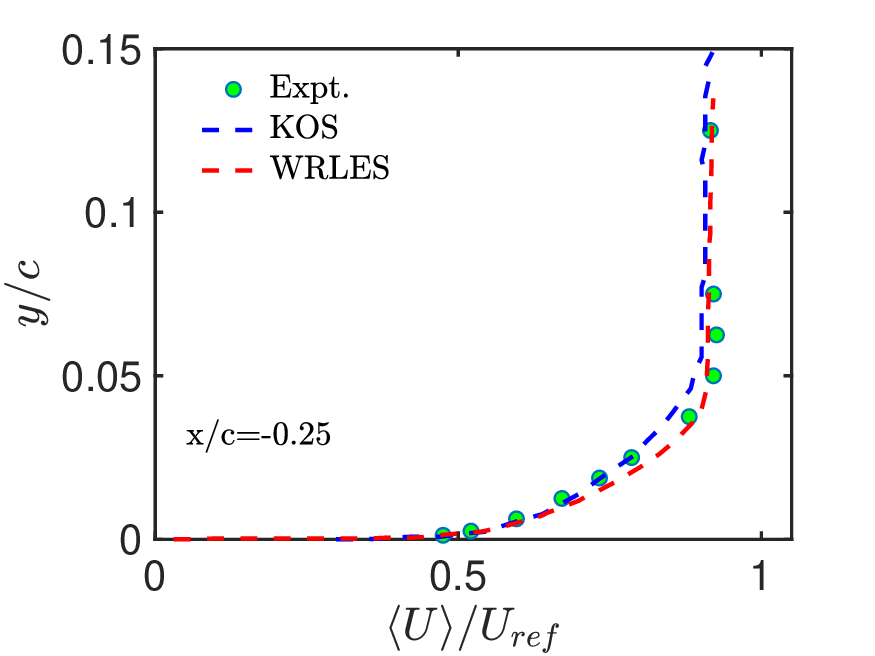}} 
			{\includegraphics[width=0.25\textwidth]{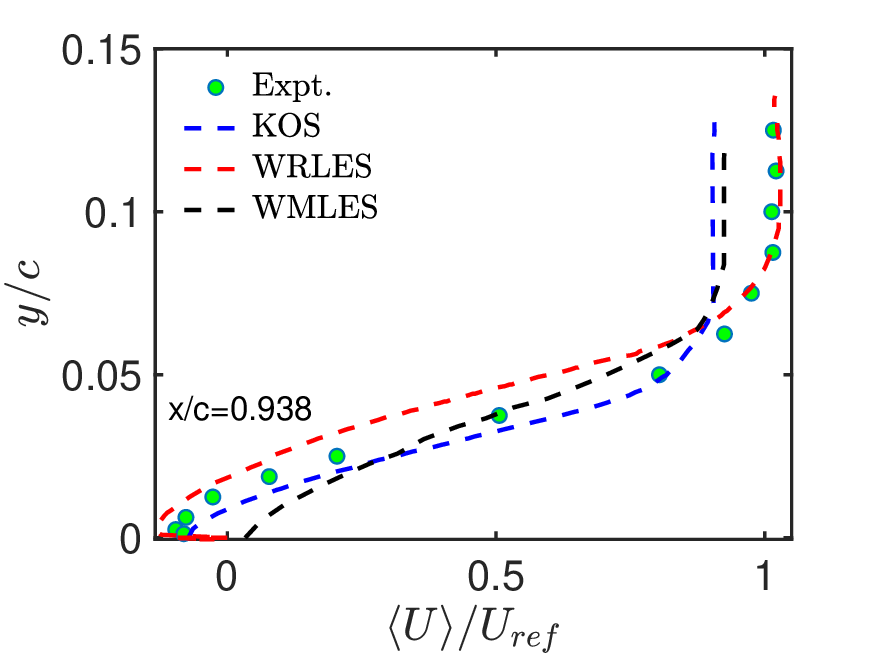}} 
			{\includegraphics[width=0.25\textwidth]{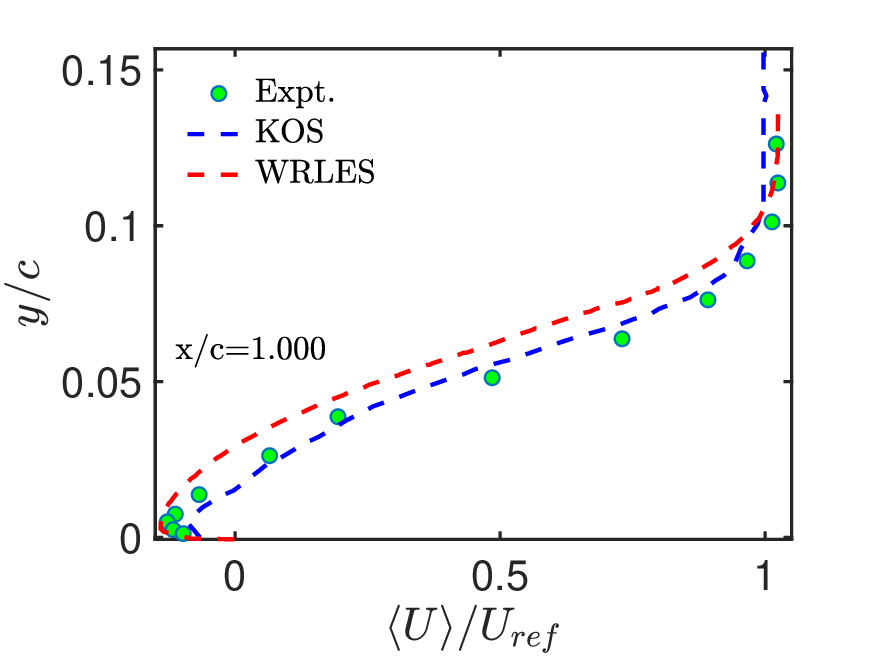}} 
			{\includegraphics[width=0.25\textwidth]{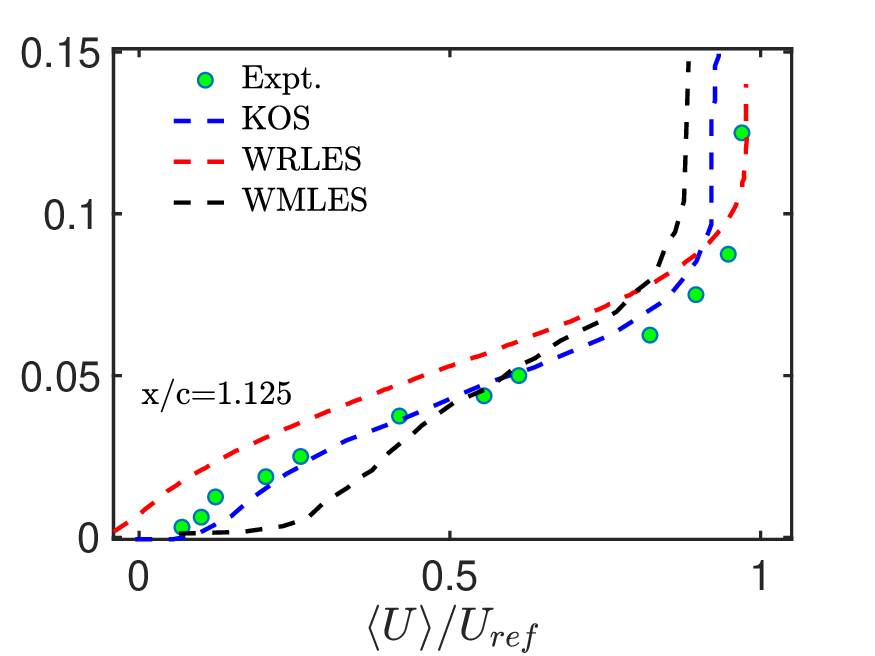}} \\
	{\includegraphics[width=0.25\textwidth]{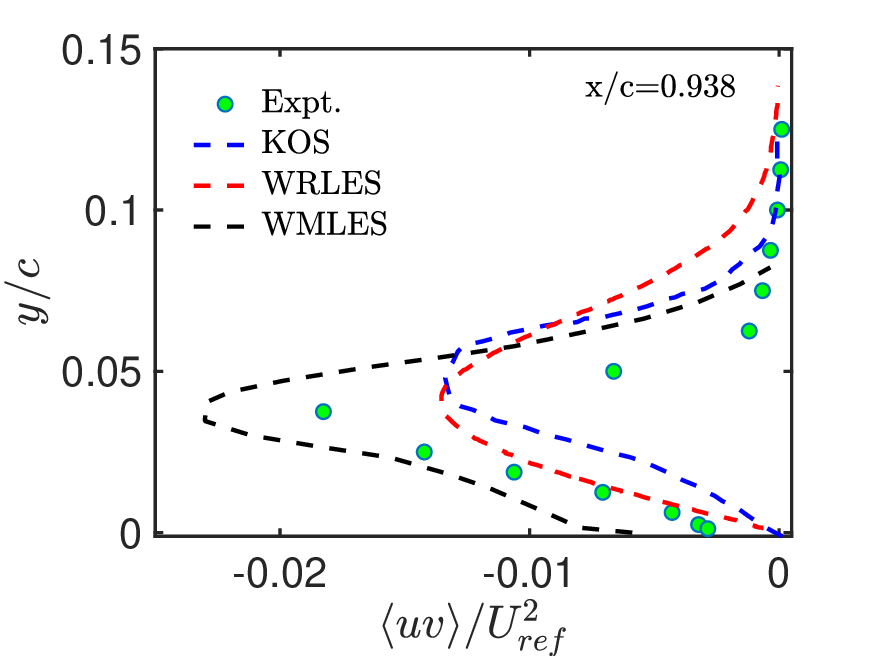}}  
			{\includegraphics[width=0.25\textwidth]{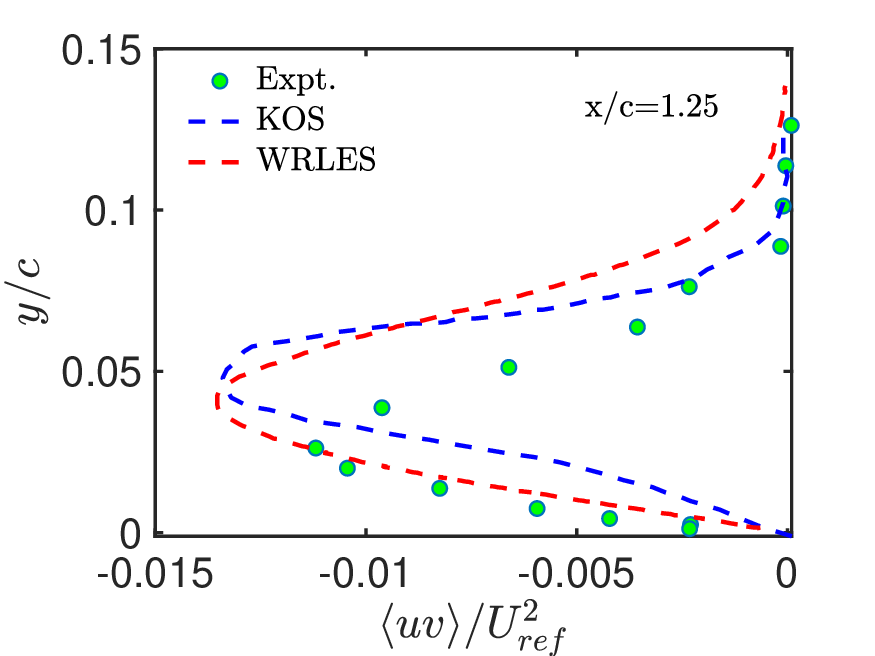}}
			{\includegraphics[width=0.25\textwidth]{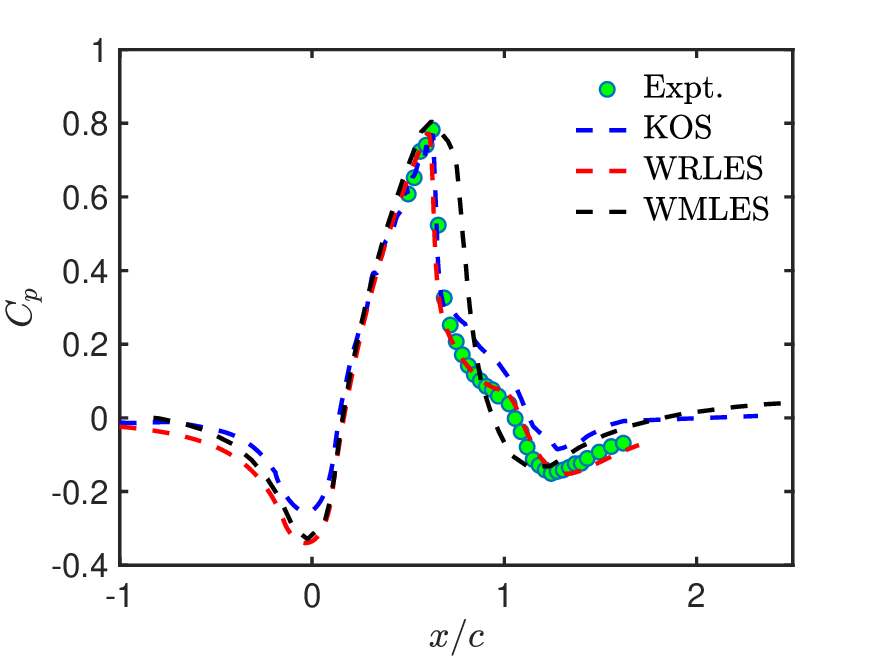}}
			{\includegraphics[width=0.25\textwidth]{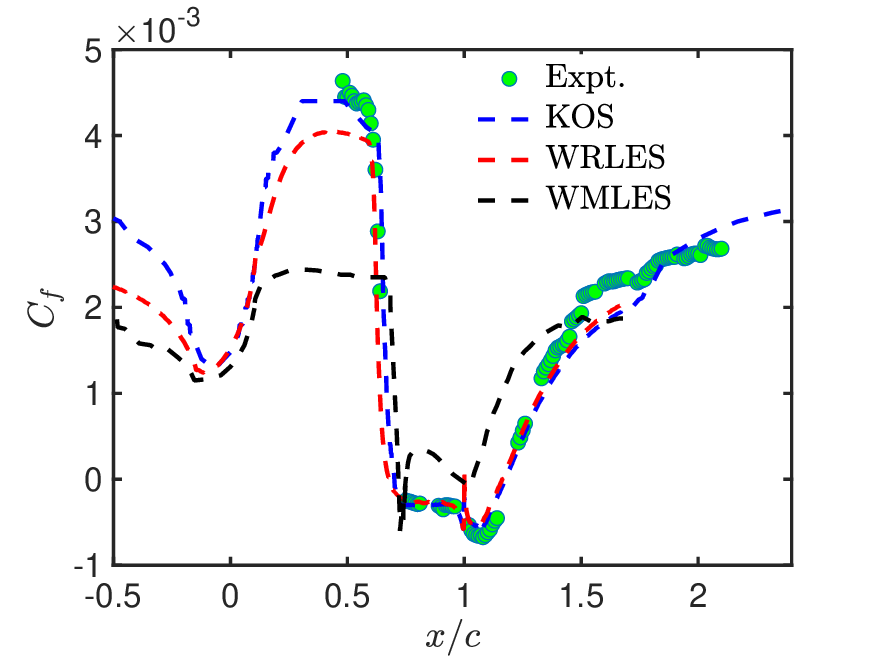}}\\
		\end{tabular}
	\end{center}
	\vspace{-.2in}
	\caption{CES-KOS vs. LES-type WRLES~\protect{\citep{UZUNMALIK-19} and WMLES~\citep{Xiang_et_all} models: 
	Profiles of the normalized streamwise velocity $\langle U\rangle/U_{ref}$, Reynolds stress $\langle uv\rangle/U^{2}_{ref}$, pressure and skin-friction coefficients at different locations.}}
	\label{fig:FCLpp8a}
\end{figure*}

Figure~\ref{fig:FCLpp8a} shows in its first row streamwise velocity profiles obtained by CES-KOS, WMLES~\citep{Xiang_et_all} and WRLES~\citep{UZUNMALIK-19}. 
It may be seen that the CES-KOS model predicts the streamwise velocity more accurately than WMLES and WRLES. 
In regard to turbulent shear stress profiles shown in the second row, we see a reasonable agreement of WMLES, WRLES, and CES-KOS with experimental data. It is of interest to note that CES-KOS and WRLES provide very similar results. 
In attached flow regions, WMLES over-predicts the turbulent shear stress. Due to its delayed reattachment point, WMLES predicts a faster separated shear layer growth and a higher maximum Reynolds stresses compared to CES-KOS. These trends have been noted in past numerical studies using alternative models~\citep{SahuDanbergetal}.

Figure~\ref{fig:FCLpp8a} also shows pressure coefficient distributions obtained by CES-KOS, WMLES, and WRLES. 
 The figure helps to illustrate and validate the accuracy of CES-KOS predictions.
The CES-KOS and WRLES models accurately predict pressure coefficient profiles due to their sufficient flow resolution and ability. In contrast, WMLES predicts a linearly increasing pressure distribution within $x/c = (0.7,1.1)$, it fails to accurately capture the separation zone. 
Furthermore, both CES-KOS and WRLES show reasonable predictions of the shock location and post-shock pressure recovery. The WRLES results agree slightly better with the experimental data downstream of the bump (between $x/c = 1.1$ and $1.3$) compared to the CES-KOS model.
Figure~\ref{fig:FCLpp8a} also shows skin-friction  coefficient distributions obtained by CES-KOS, WMLES, and WRLES. 
Evidently, WMLES significantly underestimates the skin-friction  coefficient in the separation region and fails to accurately represent the post-separation flow physics. The predictions of CES-KOS and WRLES are very similar, with the exception that CES-KOS better agrees with the experimental data in the $C_{f}$  plateau region upstream of separation. Overall, CES-KOS provides the most accurate predictions.

\section*{SUMMARY}
\label{sec:sum}

The paper reports the advantages of CES methods, i.e. partially resolving simulation methods based on strict mathematics. 
Corresponding minimal error simulation methods include an essential mechanism that is missing in popular hybrid RANS-LES methods: the model integrates physics, it can dynamically respond to the actual amount of flow resolution, which is the essential mechanism to ensure a functional RANS-LES swing. This theoretical advantage relates to functionality features different from  popular hybrid RANS-LES methods. WMLES and DES methods usually applied depend on a variety of simulations settings which are usually determined to maximize the simulation performance. In contrast CES methods do only depend on the details of the turbulence model applied but not on adjustable settings of the hybridization. In addition CES methods are computationally much more efficient than usually applied hybrid RANS-LES and WRLES. 

Based on these advantages, CES methods were found to perform significantly better than DES and WMLES simulations and at least as good (or even better) than WRLES. The latter was shown for three complex flow applications: periodic hill flows at high Reynolds number~\citep{PF20}, the NASA wall-mounted hump flow~\citep{Fluids24-CES}, and the Bachalo \& Johnson axisymmetric transonic bump flow~\citep{ApplSci24-CES}. An interesting overall observation of these applications was the fact that CES predictions are well balanced. This means the usual problem of DES, WMLES, and WRLES to perform well (not well) in regard to specific flow features (like velocities distributions, pressure and skin-friction distributions) was not observed in regard to CES predictions. Based on the stable functioning of CES methods, we discussed the asymptotic flow structure of the three flows considered at extreme Reynolds numbers.

We note that the use of CES methods as resolving LES is highly attractive to avoid the problem to involve the filter width as artificial (possibly unphysical) length scale. Similarly, the use of CES methods in almost RANS mode will relate to significant advantages because of the stable inclusion of unsteady turbulence.

\bibliographystyle{tsfp}


\end{document}